\documentclass[aps,prb,nobibnotes,twocolumn,superscriptaddress,bibliography]{revtex4-2}

\usepackage{amsfonts}
\usepackage{mathrsfs}
\usepackage{amsmath}
\usepackage{color}
\usepackage{graphicx}
\usepackage{bm}
\usepackage{amssymb}
\usepackage{xspace}
\usepackage{epstopdf}
\usepackage{dcolumn}
\usepackage{longtable}
\usepackage{multirow}
\usepackage{float}
\usepackage{comment}
\usepackage{makecell}
\usepackage{multirow}

\usepackage[colorlinks=true, letterpaper=true, pdfstartview=FitV, linkcolor=blue, citecolor=blue, urlcolor=blue]{hyperref}

\makeatother
\newcommand{\ket}[1]{|#1\rangle}



\begin{document}

\title{Rotation topological states: theory and material realization}

\author{Chun-Xue Liu}
\affiliation{Key Lab of advanced optoelectronic quantum architecture and measurement (MOE), Beijing Key Laboratory of Quantum Matter State Control and Ultra-Precision Measurement Technology, and School of Physics, Beijing Institute of Technology, Beijing 100081, China}

\author{Yilin Han}
\email{yilinhan@bit.edu.cn}
\affiliation{Key Lab of advanced optoelectronic quantum architecture and measurement (MOE), Beijing Key Laboratory of Quantum Matter State Control and Ultra-Precision Measurement Technology, and School of Physics, Beijing Institute of Technology, Beijing 100081, China}

\author{Runze Li}
\affiliation{Key Lab of advanced optoelectronic quantum architecture and measurement (MOE), Beijing Key Laboratory of Quantum Matter State Control and Ultra-Precision Measurement Technology, and School of Physics, Beijing Institute of Technology, Beijing 100081, China}

\author{Yulong Liu}
\email{yulongliu@bit.edu.cn}
\affiliation{Key Lab of advanced optoelectronic quantum architecture and measurement (MOE), Beijing Key Laboratory of Quantum Matter State Control and Ultra-Precision Measurement Technology, and School of Physics, Beijing Institute of Technology, Beijing 100081, China}

\author{Zhi-Ming Yu}
\affiliation{Key Lab of advanced optoelectronic quantum architecture and measurement (MOE), Beijing Key Laboratory of Quantum Matter State Control and Ultra-Precision Measurement Technology, and School of Physics, Beijing Institute of Technology, Beijing 100081, China}

\begin{abstract}
The conventional characterization of topological materials relies on topological invariants calculated from the entire set of occupied bands.
However, when a system possesses rotational symmetry, the occupied Hilbert space can be decomposed into multiple subspaces labeled by distinct rotation eigenvalues. We show that this decomposition reveals hidden topological states characterized by a novel $\mathbb{Z}_2^n$ topological invariant, where $n$ is the number of subspaces, while the conventional $\mathbb{Z}_2$ invariant may fail to detect the topology hidden in the rotation subspaces.
Remarkably, time-reversal symmetry pairs conjugate rotation eigenvalues and guarantees that the two subspaces have the same $\mathbb{Z}_2$ invariants, making the topology always hidden from the conventional global invariant.
We formulate the theory of rotation-subspace topology and demonstrate its material realization in bulk CsCl.
Using first-principles calculations and symmetry analysis, we show that bulk CsCl, which is diagnosed as topologically trivial by the conventional approach, features a nontrivial $\mathbb{Z}_2^3$ invariant along the $\Gamma$-R path and a nontrivial $\mathbb{Z}_2^4$ invariant along the $\Gamma$-Z and M-R paths, leading to double Weyl points on the (111) and (001) surfaces, respectively.
The subspace $\mathbb{Z}_2^n$ invariant proposed here serves as a necessary refinement for symmetry-protected topological phases and will facilitate the identification of a large class of topological states overlooked by existing diagnostics.
\end{abstract}

\maketitle
\section{Introduction}
Topological insulators (TIs) represent one of the most significant discoveries in condensed matter physics over the past two decades~\cite{RevModPhys.82.3045,RevModPhys.83.1057,RevModPhys.88.035005,Lv2019,science.1133734,PhysRevLett.98.106803,Zhang2009,Moore2010,Zhao2020,Shumiya2022,Han2024,PhysRevLett.133.176602,5yxp-djy9}.
A unique feature of TIs is the bulk-boundary correspondence: TIs are characterized by a nontrivial topological invariant defined in the bulk and host topological boundary states on their boundaries. There exist many kinds of topological invariants, such as the $\mathbb{Z}$ invariant in Chern insulators and the $\mathbb{Z}_2$ invariant in quantum spin Hall insulators~\cite{Bernevig2013,Vanderbilt2018,Konig2007,PhysRevLett.96.106802}.
The introduction of crystalline symmetries has opened the door to a refined perspective~\cite{PhysRevLett.106.106802,PhysRevB.86.115112,Ando2015,PhysRevX.7.041069,PhysRevX.8.031069,Science}.
In particular, topological quantum chemistry and symmetry indicators provide a systematic approach to classifying topological invariants enabled by symmetry operations~\cite{Bradlyn2017,Song2018,Tang2019Efficient,Elcoro2021Magnetic}.
Guided by these theories, various topological states have been unveiled~\cite{Vergniory2019,Zhang2019,Tang2019Comprehensive}.
In these frameworks, the full Hilbert space of occupied states is used to analyze topological invariants~\cite{Vergniory2019,Zhang2019}.

\begin{figure}[b]
	{\includegraphics[clip,width=8.5cm]{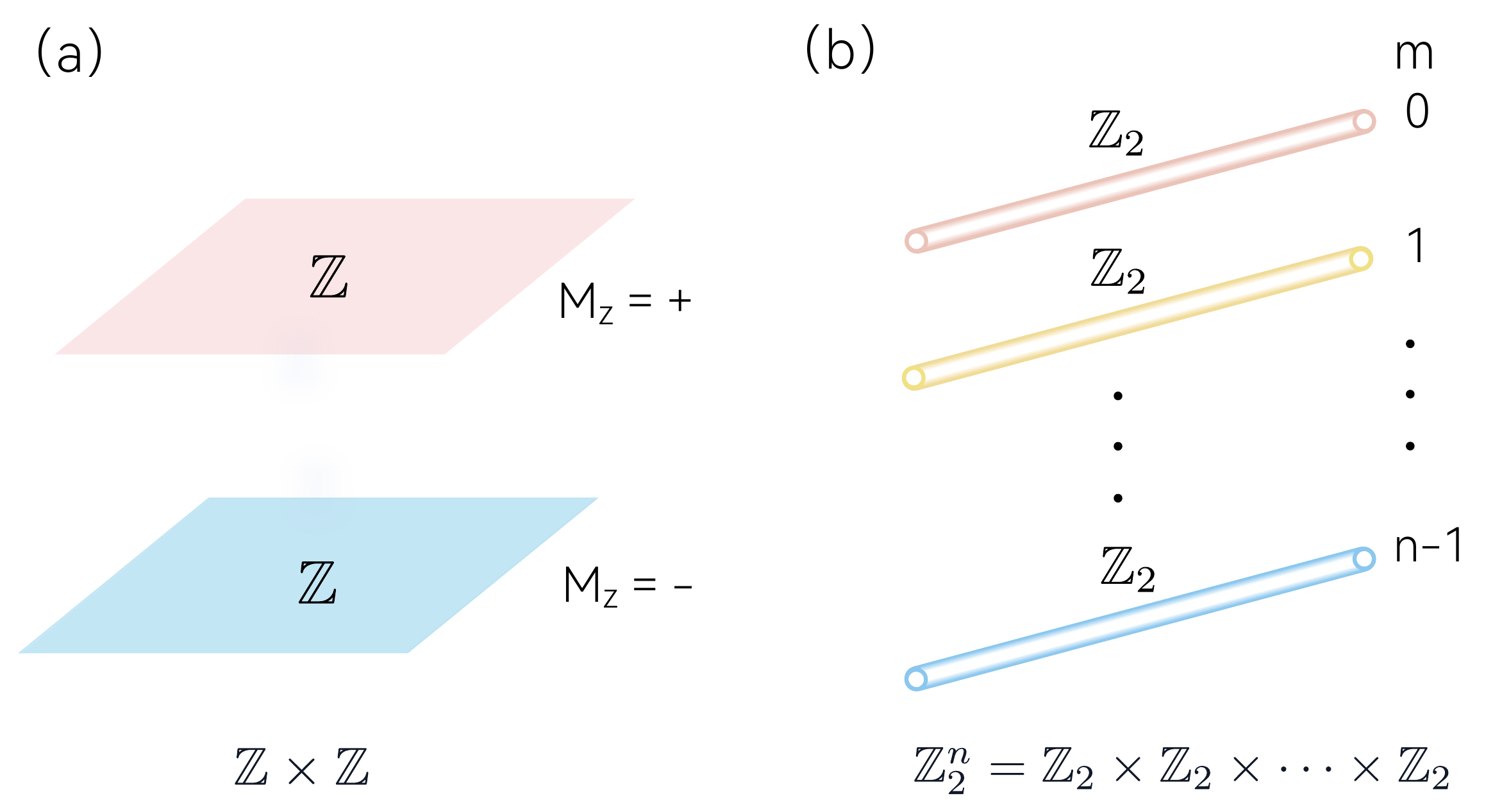}}
	\caption{\label{Fig1} (a) With mirror symmetry, the Hilbert space of a system can be divided into $M_z=+$ and $M_z=-$ subspaces. Each subspace can independently carry a $\mathbb{Z}$-valued Chern number, giving rise to an overall $\mathbb{Z}\times\mathbb{Z}$ classification. (b) Schematic illustration of rotation-subspace topology. With $n$-fold rotation symmetry, the Hilbert space of a system can be divided into multiple rotation subspaces, labeled by $m=0,1,\ldots,n-1$. Each subspace can independently carry a $\mathbb{Z}_2$ topological invariant, resulting in an overall $\mathbb{Z}_2^n$ classification.
 }
\end{figure}

The mirror Chern insulator \cite{PhysRevB.78.045426,s9mm-5662} and the recently proposed mirror real Chern insulator \cite{gong2024hidden,PhysRevB.109.195101} provide a new possibility for defining topology.
Mirror symmetry decomposes the Hilbert space into even- and odd-parity subspaces, each of which can independently carry a (real) Chern number, giving rise to the mirror (real) Chern insulator, as illustrated in Fig.~\hyperref[Fig1]{\ref*{Fig1}(a)}.
This suggests that crystalline symmetry enables a subspace decomposition of the Hilbert space \cite{p3xh-f5bx}, and each subspace can independently host a topological phase.

Besides mirror symmetry, rotation symmetries are another common class of crystalline symmetry operations and can decompose the Bloch states along one-dimensional (1D) symmetric paths into  multiple invariant subspaces labeled by rotation eigenvalues.
Generally,  1D TIs are characterized by a  $\mathbb{Z}_2$ invariant, which is evaluated from the eigenvalues of  spatial inversion (${\cal P}$) of the valence bands at two ${\cal P}$-invariant points \cite{PhysRevLett.62.2747,PhysRevB.76.045302,PhysRevB.83.245132,PhysRevB.89.155114}.
An important  question naturally arises: Can  rotational symmetries enable new topological states, similar to mirror symmetry?

In this work, we develop the theory of rotation-subspace topology and demonstrate its  material realization in bulk CsCl.
We show that, in a 1D system or along a 1D high-symmetry path in a higher-dimensional system, each rotation subspace  can feature an individual $\mathbb{Z}_2$ invariant, leading to a novel $\mathbb{Z}_2^n$ topological invariant, where $n$ is the number of the subspaces, as illustrated in Fig.~\hyperref[Fig1]{\ref*{Fig1}(b)}.
The conventional  $\mathbb{Z}_2$ invariant is the sum of these subspace invariants modulo $2$.
When an odd number of subspaces are topologically nontrivial, the conventional invariant is also nontrivial.
In sharp contrast, when an even number of subspaces are topologically nontrivial, the conventional invariant is trivial.
This means that in such case, the rotation topology is hidden and can not be diagnosed by  conventional methods.
More importantly, time-reversal symmetry (${\cal T}$)  pairs conjugate rotation eigenvalues and guarantees that these two subspaces have the same $\mathbb{Z}_2$ invariants, making the topology of such systems always hidden from the global invariant.
We further demonstrate our idea by identifying CsCl as an ideal material candidate for rotation topological states, which exhibits a nontrivial $\mathbb{Z}_2^3$  invariant along the $\Gamma$-R path and nontrivial $\mathbb{Z}_2^4$  invariant along the $\Gamma$-Z and M-R paths.
Our findings establish rotation-subspace topology as an important   refinement of conventional topological classification and open new directions for discovering topological materials hidden by conventional global invariants.

\section{Rotation subspace topology}
\subsection{Subspace decomposition and  $\mathbb{Z}_2^n$ invariant}
Consider a  system with an $n$-fold rotational symmetry ${\cal C}_n$ along a specific axis.
This system can be either a 1D, 2D or 3D material. For 2D or 3D systems, we focus on 1D high-symmetry paths in the Brillouin zone that exhibit ${\cal C}_n$ symmetry.
Obviously, if a 1D high-symmetry path has nontrivial topology, it will  lead to  topological  states within the bulk band gap  at the boundary normal to the high-symmetry path, as guaranteed by the bulk-boundary correspondence.

Along a high-symmetry path that is invariant under ${\cal C}_n$, the Hamiltonian of the system  commutes with the rotation operator: $[\mathcal{H}(\mathbf{k}), {\cal C}_n] = 0$. Therefore, the Bloch states of the system  can  be  simultaneous eigenstates of $\mathcal{H}(\mathbf{k})$ and ${\cal C}_n$, with eigenvalues
\begin{equation}
{\cal C}_n \ket{u_m(\mathbf{k})} = e^{i2\pi m/n} \ket{u_m(\mathbf{k})},
\end{equation}
for spinless systems. Here, $\ket{u_m(\mathbf{k})} $ denotes a Bloch eigenstate and $m = 0,1,\dots,n-1$.
The eigenvalue $m$ constitutes a conserved quantum number along the symmetric path, and the Bloch Hamiltonian can be  decomposed into a direct sum of  sub-Hamiltonian on the invariant subspaces:
\begin{equation}
\mathcal{H}(\mathbf{k}) = \bigoplus_{m=0}^{n-1} \mathcal{H}_m(\mathbf{k}).
\end{equation}
Similarly, the entire set of occupied valence bands can also be decomposed into a direct sum of bands in the  rotation subspaces.
Each subspace defined by  $\mathcal{H}_m(\mathbf{k})$ evolves independently as a function of $\mathbf{k}$ along the symmetric path, forming a well-defined 1D submanifold within the Brillouin zone (BZ). Note that each subspace can be viewed as an independent  system with its own band structure and topological character.

In the presence of additional ${\cal P}$ symmetry, we can define a $\mathbb{Z}_2$ invariant for each  1D subspace $\mathcal{H}_m$.
For the $m$-th subspace, the  $\mathbb{Z}_2$ invariant  is evaluated from the ${\cal P}$ eigenvalues of the valence bands in that subspace, which has rotation eigenvalue  $e^{i2\pi m/n}$,
\begin{equation}
(-1)^{\nu_m}= (-1)^{n_m^{\mathbf{k}=0}+n_m^{\mathbf{k}=\pi}},
\end{equation}
where  $\nu_m$ is the $\mathbb{Z}_2$ invariant of  $m$-th subspace, and $n_m^{\mathbf{k}=0}$ and $n_m^{\mathbf{k}=\pi}$ are the numbers of valence states with negative ${\cal P}$ eigenvalues in the $m$-th subspace at the two ${\cal P}$-invariant points \cite{Bernevig2013,Vanderbilt2018}.
Since each subspace gives an independent $\mathbb{Z}_2$ invariant,  the complete topological classification of the 1D (sub)system with ${\cal C}_n$ symmetry is $\mathbb{Z}_2^n$ rather than a single $\mathbb{Z}_2$.

Because  the set of all occupied states is the direct sum of all subspaces, the conventional global topological invariant can be  obtained from the subspace $\mathbb{Z}_2$ invariants:
\begin{equation}
(-1)^{\nu_{\rm global}} = \prod_m (-1)^{\nu_m},
\end{equation}
or equivalently,
\begin{equation}
\nu_{\rm global} = \left( \sum_m \nu_m \right) \bmod 2.
\label{eq:multiplicative}
\end{equation}
Equation~(\ref{eq:multiplicative}) reveals a fundamental limitation of the conventional approach: the global $\mathbb{Z}_2$ invariant detects only the parity of the number of nontrivial subspaces. When an odd number of subspaces carry nontrivial topology, the global invariant correctly signals a nontrivial phase. However, when an even number of subspaces are nontrivial, one always obtains $\nu_{\rm global} = 0$, and the system is incorrectly classified as topologically trivial by any diagnostic based solely on the global invariant.

\subsection{${\cal T}$ induced inevitable pairing}
Time-reversal symmetry ${\cal T}$  introduces a further constraint that makes this limitation systematic rather than accidental. For a ${\cal C}_n$ rotation, the eigenvalues $e^{i2\pi m/n}$ and $e^{i2\pi (n-m)/n}$ are  complex conjugates for $m \neq 0, n/2$. ${\cal T}$ maps an eigenstate with eigenvalue $e^{i2\pi m/n}$ to that  with eigenvalue $e^{i2\pi (n-m)/n}$. This establishes a Kramers-like pairing between the subspaces $\mathcal{H}_m$ and $\mathcal{H}_{n-m}$.
More importantly, since ${\cal T}$ preserves the parity eigenvalue at ${\cal P}$-invariant points \cite{PhysRevB.76.045302}, the $\mathbb{Z}_2$ invariants  of the paired subspaces are forced to be equal,
\begin{equation}
\nu_m = \nu_{n-m}.
\label{eq:TRS_pairing}
\end{equation}
Therefore, in the presence of ${\cal T}$ symmetry,  nontrivial rotation  subspaces necessarily appear in pairs for $m \neq 0, n/2$. Thus, the nontrivial topology of the $m$-th ($m \neq 0, n/2$) rotation  subspace is always missed by the conventional diagnosis, because   Eq.~(\ref{eq:multiplicative}) guarantees $\nu_{\rm global} = 0$.
This is not accidental, but rather a generic situation for ${\cal T}$-protected topological phases with rotational symmetry.
It should be mentioned that while the above analysis is focused on  spinless systems, the extension to spinful systems with spin-orbit coupling (SOC)  is straightforward.

\subsection{Paired  bulk-boundary correspondence}
The standard bulk-boundary correspondence naturally extends to the subspace level. Each subspace $\mathcal{H}_m$ with $\nu_m = 1$ contributes a protected boundary state at the boundary normal to the symmetric path. In a 3D crystal, the topological  boundary states appear at a high-symmetry ${\cal C}_n$-invariant  point in the surface BZ.
The total number of boundary states equals the number of nontrivial subspaces along the high-symmetry path.
Interestingly, for a system with rotation symmetry and   nontrivial  ${\cal T}$-paired  rotation  subspaces, a symmetry-protected doubly degenerate boundary states are expected.
This indicates that two surface states will exist in the surface BZ of a 3D system, and these two surface states will exhibit a band degeneracy, which   is  protected by ${\cal T}$ and ${\cal C}_n$.

In the following, we demonstrate our idea  through first-principles calculations on CsCl, which, despite its global triviality, hosts rich subspace topology protected by ${\cal C}_3$ and ${\cal C}_4$ symmetries.

\section{Subspace topology in CsCl}
\subsection{Computational details}
The first-principles calculations were performed within the framework of density functional theory (DFT) using the Vienna \emph{Ab initio} Simulation Package (VASP) \cite{PhysRev.136.B864,PhysRev.140.A1133,PhysRevB.54.11169}. The exchange-correlation functional was modeled by the generalized gradient approximation (GGA) with the Perdew-Burke-Ernzerhof (PBE) realization \cite{PhysRevLett.77.3865}. The plane-wave cutoff energy was set to 450~eV. For Brillouin-zone sampling, a $\Gamma$-centered $6\times 6\times 6$ mesh, and a Gaussian smearing of 0.01~eV was used for electronic occupations \cite{PhysRevB.13.5188,PhysRevB.40.3616}. The crystal structure was optimized until the ionic forces converged to below 0.01~eV/\AA. After relaxation, the lattice parameter of CsCl was determined to be $a=b=c=4.143$~\AA.

To calculate the topological properties, Wannier functions for CsCl were constructed from DFT bands using Wannier90 ~\cite{PhysRevB.56.12847,PhysRevB.65.035109,MOSTOFI2008685}. Based on the constructed tight-binding model, the surface states were calculated using WannierTools \cite{WU2018405}. The irreducible representations of electronic states were calculated using the Irvsp package \cite{GAO2021107760}.

\begin{figure}[t]
	{\includegraphics[clip,width=8.5cm]{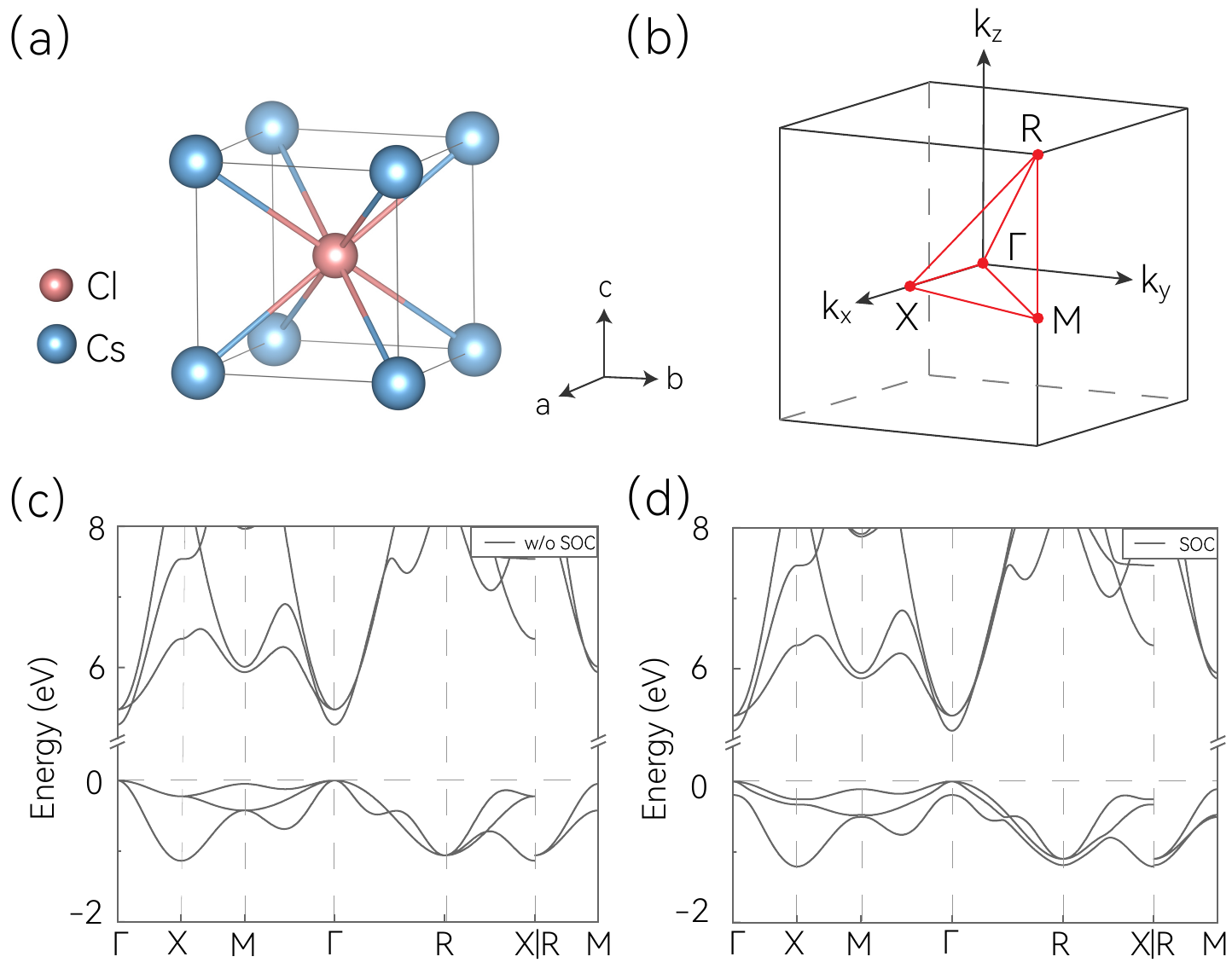}}
	\caption{\label{Fig2} (a) Crystalline structure of CsCl. (b) Bulk Brillouin zone (BZ) and the relevant high-symmetry points. (c,d) Bulk band structures of CsCl calculated (c) without SOC and (d) with SOC.}
\end{figure}

\subsection{Subspace topology without SOC}
CsCl is an ideal platform for demonstrating rotation-subspace topology, because it is a realistic material diagnosed as a trivial band insulator and its simple cubic structure provides clean ${\cal C}_3$-  and ${\cal C}_4$-symmetric paths.

The crystal structure of CsCl is shown in Fig.~\hyperref[Fig2]{\ref*{Fig2}(a)}. It belongs to the cubic crystal system with space group $Pm\bar{3}m$ (No.~221), possessing ${\cal P}$, multiple rotational symmetries and ${\cal T}$.
Figure~\hyperref[Fig2]{\ref*{Fig2}(b)} shows the BZ with the relevant high-symmetry paths. Two symmetry-inequivalent types of rotational symmetry are relevant here: (i) the $\Gamma$-R path along the [111] direction preserves ${\cal C}_{3,111}$ symmetry; and (ii) the $\Gamma$-Z and M-R paths along the [001] direction preserve ${\cal C}_{4z}$ symmetry.

The electronic band structures of CsCl without and with SOC are respectively shown in Figs.~\hyperref[Fig2]{\ref*{Fig2}(c,d)},  where a large band gap of $\sim 5$ eV can be observed. Since the low-energy bands remain almost unchanged upon the inclusion of spin-orbit coupling, SOC is neglected in all subsequent topological analyses. The parity eigenvalues of the entire set of occupied bands and the corresponding conventional global $\mathbb{Z}_2$ invariant ($\nu_{\rm global}$) along the $\Gamma$-R, $\Gamma$-Z and M-R paths are listed in Tables \ref{tab:TABLE I} and \ref{tab:TABLE III}, showing that  CsCl is trivial under conventional diagnosis.

\subsubsection{$\Gamma$-R path: hidden topology under ${\cal C}_3$ symmetry}
The occupied bands along $\Gamma$-R can be decomposed into three invariant subspaces labeled by $m=0,1,2$, corresponding to subspaces with  ${\cal C}_3$ eigenvalues $e^{i 2\pi m/3}$. Table~\ref{tab:Table II} presents the parity analysis within each subspace.
The $m=0$ subspace gives $\nu_{0} = 0$, while the  $m=1$ and $m=2$ subspaces, which are paired by ${\cal T}$, each give  $\nu_{1} = \nu_{2} = 1$, resulting in a nontrivial topological $\mathbb{Z}_2^3$ $(=\mathbb{Z}_2\times \mathbb{Z}_2\times \mathbb{Z}_2)$  invariant:  $(\nu_{0},\nu_{1},\nu_{2})=(0,1,1)$.
This means that the $m=0$ subspace  is trivial, whereas both the $m=1$ and $m=2$ subspaces are nontrivial.
With the two  ${\cal T}$-paired nontrivial subspaces, the global invariant is $\nu_{\rm global} = (1+1) \bmod 2 = 0$, masking the rich subspace topology. This is precisely the case  described in Eq.~(\ref{eq:TRS_pairing}): the conventional $\mathbb{Z}_2$ analysis is insufficient for the systems with two or multiple symmetry-decoupled subspaces.

\begin{figure}[t]
	{\includegraphics[clip,width=8.5 cm]{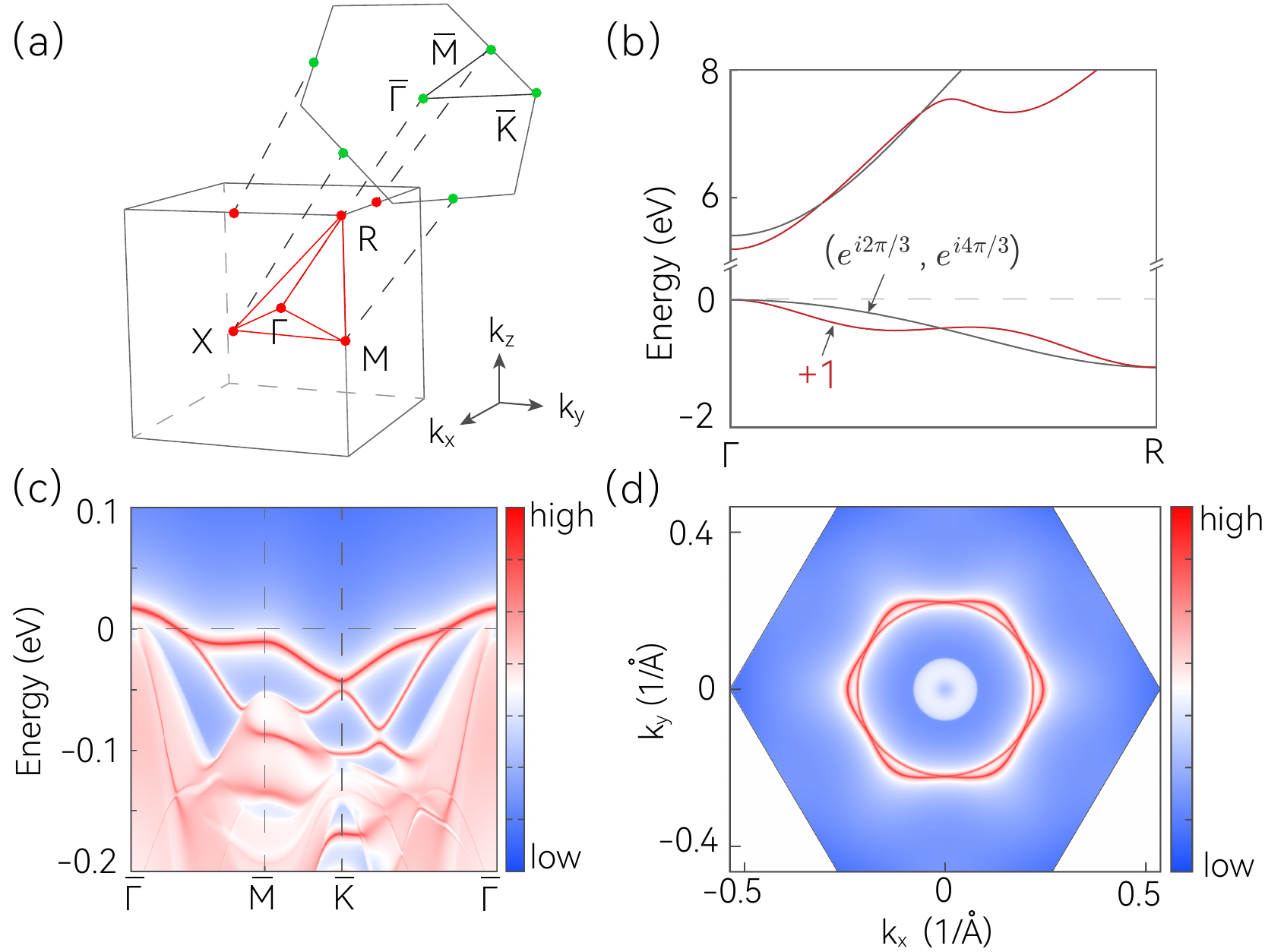}}
	\caption{\label{Fig3} (a) Bulk BZ and the (111)-surface BZ. (b) Bulk band structure along the $\Gamma$-R high-symmetry line without SOC. The ${\cal C}_3$ rotation eigenvalues of the relevant bands are indicated. The gray bands denote a pair of degenerate bands. (c) Surface spectrum on the (111) surface. (d) Constant-energy slice of the (111) surface spectrum at $E = 0$ eV, as marked in (c). }
\end{figure}

\begin{table}[b]
    \centering
    \caption{Parity analysis of all occupied bands for  $\Gamma$-R path.}
    \label{tab:TABLE I}
    \begin{ruledtabular}
        \begin{tabular}{ccc}
            whole system & $\Gamma$ & R \\
            \hline
            $n_+$  & 2 & 4 \\
            $n_-$  & 6 & 4 \\
            $\nu$ & \multicolumn{2}{c}{$0$} \\
        \end{tabular}
    \end{ruledtabular}
\end{table}

\begin{table}[t]
    \centering
    \caption{Parity analysis of each ${\cal C}_3$ subspace for $\Gamma$-R path. The $m=1,2$ subspaces exhibit odd band inversions ($\nu_1=\nu_2=1$), and are topologically nontrivial.}
    \label{tab:Table II}
    \begin{ruledtabular}
        \begin{tabular}{ccccc}
            \multirow{2}{*}{Subspace} & \multicolumn{2}{c}{$m=0$} & \multicolumn{2}{c}{$m=1,2$} \\
            & $\Gamma$ & R & $\Gamma$ & R \\ \hline
            $n_+$ & 2 & 2 & 0 & 1 \\
            $n_-$ & 2 & 2 & 2 & 1 \\
            $\nu$ & \multicolumn{2}{c}{$0$} & \multicolumn{2}{c}{$1$} \\
        \end{tabular}
    \end{ruledtabular}
\end{table}

The subspace bulk-boundary correspondence predicts that each nontrivial subspace along $\Gamma$-R  contributes one protected boundary  state at the  $\bar{\Gamma}$ point in the (111) surface BZ [see Fig.~\hyperref[Fig3]{\ref*{Fig3}(c)}]. The $m=1$ and $m=2$ subspaces thus together produce a twofold-degenerate state, forming a Weyl point protected by  ${\cal T}$ and ${\cal C}_3$ symmetry.
Interestingly, since the boundary states at the $\bar{\Gamma}$ point originate from the $m=1$ and $m=2$ subspaces, their  ${\cal C}_3$ eigenvalues must be $e^{i 2\pi/3}$ and $e^{i 4\pi/3}$. Using these two states as a basis, we can construct the $k\cdot p$ effective model based on the symmetry of  the $\bar{\Gamma}$ point, of which the generators can be chosen as  ${\cal C}_3$, $\mathcal{M}_{110}$, and ${\cal T}$.
The matrix representations of the generators under  the basis of $\{|e^{i 2\pi/3}\rangle, |e^{i 4\pi/3}\rangle\}$ can be expressed as
\begin{equation}
\begin{aligned}
\mathcal{C}_3
&=
\left[
\begin{array}{cc}
e^{i2\pi/3} & 0\\
0 & e^{i4\pi/3}
\end{array}
\right], \\
\mathcal{M}_{110}
&=
\left[
\begin{array}{cc}
0 & 1\\
1 & 0
\end{array}
\right],
\quad
\mathcal{T}
=
\left[
\begin{array}{cc}
0 & 1\\
1 & 0
\end{array}
\right]\mathcal{K}.
\end{aligned}
\end{equation}
with ${\cal K}$ the  complex conjugation. The surface Hamiltonian ${\cal H}_{\rm 111}$ is required to be invariant under the symmetry transformations, i.e.,
\begin{eqnarray}
{\cal C}_3 {\cal H}_{\rm 111} (\mathbf{k}){\cal C}_3^{-1}&=&{\cal H}_{\rm 111} ({\cal C}_3 \mathbf{k}), \\
\mathcal{M}_{110}\mathcal{H}_{111}(\mathbf{k})\mathcal{M}_{110}^{-1}
&=&
\mathcal{H}_{111}(\mathcal{M}_{110}\mathbf{k}),
\\
{\cal T}{\cal H}_{\rm 111} (\mathbf{k}){\cal T}^{-1}&=&{\cal H}_{\rm 111} (-\mathbf{k}),
\end{eqnarray}
where $\mathbf{k}$ is measured from the $\bar{\Gamma}$ point. A straightforward calculation up to second order in $\mathbf{k}$ gives \cite{ZHANG2023108784}
\begin{equation}
{\cal H}_{111}({\bf k})
=c_1 k^2 +
\begin{pmatrix}
0 & c_2 k_+^2 \\
c_2 k_-^2 & 0
\end{pmatrix},
\end{equation}
where $k_\pm=k_x\pm i k_y$,   $k^2=k_x^2+k_y^2$,  $c_1$ and $c_2$ are real parameters.
The energy spectrum of this point  is $E_\pm=c_1 k^2\pm|c_2|k^2$, revealing a quadratic splitting.
Since $k_+^2=k^2e^{2i\phi}$ with $\phi=\arg (k_x+i k_y)$ carries a double phase winding around $\bar{\Gamma}$, this quadratic surface point is identified as a double Weyl point (DWP) \cite{YU2022375}, which has  quadratic energy splitting.

To verify this novel subspace bulk-boundary correspondence, we calculated the surface states on the (111) surface, which preserves ${\cal C}_3$ symmetry. The surface spectrum of the (111) surface is plotted in Fig.~\hyperref[Fig3]{\ref*{Fig3}(c)}. A clear twofold degeneracy with  quadratic splitting appears at the $\bar{\Gamma}$ point, consistent with the prediction of two topological boundary  states from the $m=1,2$ nontrivial subspaces.
Breaking ${\cal C}_3$  symmetry  would lift this degeneracy and gap out the surface  states, providing an experimental signature of their topological origin.

\subsubsection{$\Gamma$-Z  and M-R paths: hidden topology under ${\cal C}_4$ symmetry}
Both the $\Gamma$-Z  and M-R paths are invariant   under ${\cal C}_{4}$ rotation along the [001] direction.
The  ${\cal C}_{4}$ operator has four eigenvalues $e^{i2\pi m/4}$ with $m=0,1,2,3$, which divide the occupied bands into four invariant subspaces.
The  subspace-resolved parity analyses of the $\Gamma$-Z  and M-R paths within each subspace are listed in Table \ref{tab:Table IV}.

For the $\Gamma$-Z path, the $m=0$ ($m=2$) subspace gives $\nu_{0(2)} =0$, indicating a trivial band structure. However, the $m=1$ and $m=3$ subspaces each yield $\nu_{1} = \nu_{3} = 1$, leading to a nontrivial $\mathbb{Z}_2^4$  invariant:  $(\nu_{0},\nu_{1},\nu_{2},\nu_{3})=(0,1,0,1)$. Similar to  the ${\cal C}_3$ case, the $m=1$ and $m=3$ subspaces are paired by ${\cal T}$ and therefore always have the same $\mathbb{Z}_2$ invariant. As a result, the  global invariant $\nu_{\rm global} = (1+1) \bmod 2 = 0$ is trivial,  despite the presence of two topologically nontrivial subspaces.

\begin{table}[b]
    \centering
    \caption{Parity analysis of all occupied bands for  $\Gamma$-Z  and M-R paths.}
    \label{tab:TABLE III}
    \begin{ruledtabular}
        \begin{tabular}{ccccc}
            whole system & $\Gamma$ & Z & M & R \\
            \hline
            $n_+$  & 2 & 4 & 2 & 4 \\
            $n_-$  & 6 & 4 & 6 & 4 \\
            $\nu$ & \multicolumn{2}{c}{$0$} & \multicolumn{2}{c}{$0$} \\
        \end{tabular}
    \end{ruledtabular}
\end{table}

\begin{table*}[t]
    \centering
    \caption{Parity analysis of each ${\cal C}_4$ subspace for  $\Gamma$-Z  and M-R paths.}
    \label{tab:Table IV}
    \begin{ruledtabular}
        \begin{tabular}{ccccccccccccc}
            \multirow{2}{*}{Subspace} & \multicolumn{4}{c}{$m=0$} & \multicolumn{4}{c}{$m=2$} & \multicolumn{4}{c}{$m=1,3$} \\
            & $\Gamma$ & Z & M & R & $\Gamma$ & Z & M & R & $\Gamma$ & Z & M & R \\ \hline
            $n_+$ & 2 & 2 & 1 & 1 & 0 & 0 & 1 & 1 & 0 & 1 & 0 & 1  \\
            $n_-$ & 2 & 2 & 1 & 1 & 0 & 0 & 1 & 1 & 2 & 1 & 2 & 1  \\
            $\nu$ & \multicolumn{2}{c}{$0$} &  \multicolumn{2}{c}{$0$} &  \multicolumn{2}{c}{$\setminus $} &  \multicolumn{2}{c}{$0$} & \multicolumn{2}{c}{$1$} & \multicolumn{2}{c}{$1$}\\
        \end{tabular}
    \end{ruledtabular}
\end{table*}

\begin{figure}[t]
	{\includegraphics[clip,width=8.6 cm]{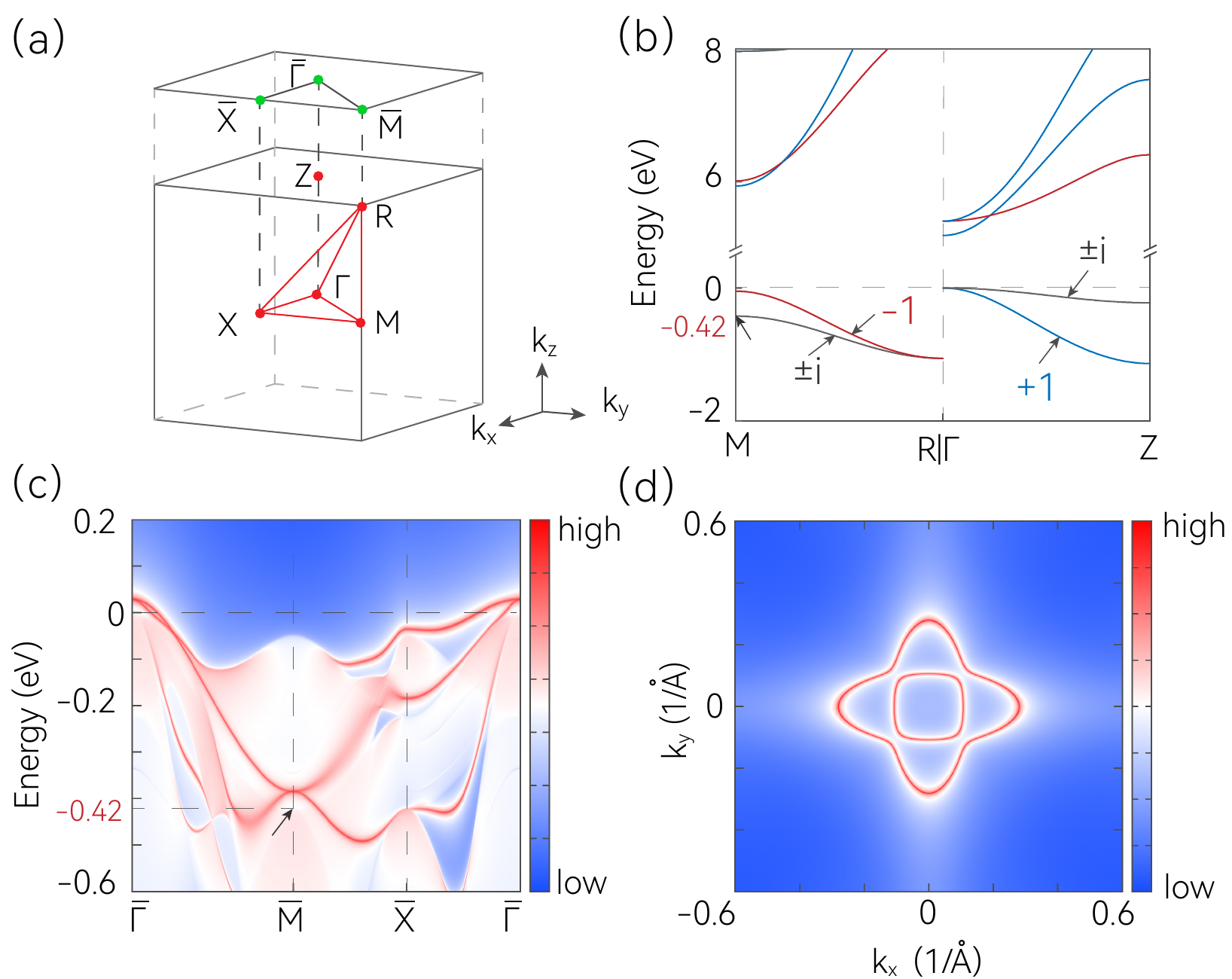}}
	\caption{
    \label{Fig4} (a) Bulk BZ and the (001)-surface BZ. (b) Bulk band structure along the M-R and $\Gamma$-Z high-symmetry paths without SOC, with the ${\cal C}_4$ rotation eigenvalues of the relevant bands indicated. The gray bands denote a pair of degenerate bands. (c) Surface spectrum on the (001) surface. (d) Constant-energy slice of the (001) surface spectrum at $E = 0$ eV. }
\end{figure}

At the $\bar{\Gamma}$ point of the (001) surface [see Fig.~\hyperref[Fig4]{\ref*{Fig4}(c)}], two topological boundary states are expected, originating from the $m=1$ and $m=3$ subspaces.
The two boundary states are degenerate, which is protected by ${\cal C}_{4}$, ${\cal T}$ and form a Weyl point at the surface bands.
We also construct the $k\cdot p$ effective model constrained by the  symmetries at $\bar{\Gamma}$, for which the generators are chosen as  $\mathcal{C}_4$, $\mathcal{M}_y$, and $\mathcal{T}$. The matrix representations of the generators under  the basis of $\{|i\rangle, |-i\rangle\}$ can be expressed as
\begin{equation}
\mathcal{C}_4 =
\left[
\begin{array}{cc}
i & 0\\
0 & -i
\end{array}
\right],
\quad
\mathcal{M}_y =
\left[
\begin{array}{cc}
0 & 1\\
1 & 0
\end{array}
\right],
\quad
\mathcal{T} =
\left[
\begin{array}{cc}
0 & 1\\
1 & 0
\end{array}
\right]\mathcal{K}.
\end{equation}
and the effective Hamiltonian expanded up to the second order is given as \cite{ZHANG2023108784}
\begin{eqnarray}
{\cal H}_{001}({\bf k}) &=&c k^2 +\alpha_1 \sigma_x \left(k_x^2-k_y^2\right) +\alpha_2 \sigma_y k_x k_y,\qquad
\end{eqnarray}
where $\sigma_{x(y)}$ is the Pauli matrix acting on the basis, and $c$, $\alpha_1$, and $\alpha_2$ are real parameters.
The energy splitting of this point  also scales as $k^2$, revealing a surface DWP.

A similar analysis shows that the M-R path also has hidden topology, as its  $\mathbb{Z}_2^4$  invariant is the same as that of the $\Gamma$-Z path: $(\nu_{0},\nu_{1},\nu_{2},\nu_{3})=(0,1,0,1)$.
Therefore, we can expect that a DWP will also appear at the $\bar{M}$ point of the (001) surface  [see Fig.~\hyperref[Fig4]{\ref*{Fig4}(c)}].
However, it should be noted that the subspace bulk-boundary correspondence predicts the existence of topological boundary states within the band gap of each subspace rather than within the global bulk gap of the entire band structure.
Along the M-R path, the subspace-resolved valence band maximum (VBM)  in $m=1$ and $m=3$ subspaces is not around the Fermi energy, and is lower than the VBM of the $m=2$ subspace.
This means that the DWP  at the $\bar{M}$ point may coexist with the bulk bands from the $m=2$  subspace in energy, but  it can be easily identified as they have different ${\cal C}_4$ eigenvalues compared with the bulk states (from $m=2$  subspace)  projected onto the $\bar{M}$ point.

Figure~\hyperref[Fig4]{\ref*{Fig4}(c)} shows the surface spectrum of the  (001) surface, in which two  DWPs at the $\bar{\Gamma}$  and  $\bar{M}$ points can be clearly observed.
Remarkably, the DWP at $\bar{\Gamma}$ lies within the global band gap, whereas that at the $\bar{M}$ point is slightly lower than the Fermi energy and coexists with the bulk bands. However, the  DWP at the $\bar{M}$ point still can be  clearly observed, as the hybridization of DWP with the bulk states is forbidden by ${\cal C}_4$ rotational symmetry.

\section{Discussions and Conclusions}
We have developed the theory of rotation-subspace topology and demonstrated its first material realization in CsCl. The main finding is that the conventional global $\mathbb{Z}_2$ topological classification is insufficient for describing the topology of  systems with rotational symmetry.
In particular, the topology in ${\cal T}$-paired subspaces is always missed by the conventional  $\mathbb{Z}_2$ invariant.
Instead, the 1D (sub)system with ${\cal C}_n$ rotation features a novel $\mathbb{Z}_2^n$ topological classification.

Our first-principles calculations of CsCl provide a concrete example. The $\Gamma$-R, $\Gamma$-Z and M-R paths are globally trivial under conventional diagnosis. However, the subspace-resolved analysis reveals nontrivial topology in the ${\cal T}$-paired subspaces. Surface state calculations on the (111) and (001) surfaces confirm the subspace bulk-boundary correspondence through the presence of DWPs  protected by ${\cal T}$ and the respective rotational symmetries.

The subspace topology proposed here  provides a more refined classification than the conventional global $\mathbb{Z}_2$ invariant, and thus offers a systematic method to recover this hidden topology  through subspace-resolved analysis.
For example, any material with rotational symmetry can be analyzed within this framework, suggesting that a large class of topological states may have been overlooked by existing diagnostics. Existing high-throughput topological materials databases, which rely on global symmetry indicators or $\mathbb{Z}_2$ invariants, may have systematically missed materials whose topology resides  in paired subspaces. A systematic re-examination of known trivial insulators with rotational symmetry is warranted.

Bulk CsCl has been experimentally synthesized. Thus, the predicted surface DWPs in  CsCl are ready for detection  by angle-resolved photoemission spectroscopy.
Moreover, since a DWP has a higher density of states than a linear WP and DWPs can appear in the surface states of several surfaces of the system, CsCl may have great potential in the field of  topological catalysis.

\begin{acknowledgements}
This work was supported by the NSF of China (Grants Nos. 125B2096, 61971044, 12234003 and 12474040).
\end{acknowledgements}

\section*{Data Availability}
All data needed to evaluate the conclusions in the paper are available within the article.
All raw data generated during the current study are available from the corresponding author on reasonable request

\bibliography{chiral_refs}
\end{document}